\documentclass[12pt]{article}

\usepackage[numbers]{natbib}
\usepackage[ruled]{algorithm2e}

\usepackage{epic}
\usepackage{eepic}
\usepackage{paralist}
\usepackage{graphicx}
\usepackage{mathrsfs}
\usepackage{tikz}
\usepackage{xcolor,colortbl}
\usepackage{wrapfig}
\usepackage{amsmath}
\allowdisplaybreaks
\usepackage{amssymb}
\usepackage{mathtools}
\allowdisplaybreaks

\newcommand{\R}{{\mathbb{R}}}
\newcommand{\Z}{{\mathbb Z}}

\newcommand{\D}{\mathcal{D}}

\newcommand{\la}{\leftarrow}
\newcommand{\ra}{\rightarrow}

\newcommand{\0}{\textbf{0}}

\newcommand{\x}{\textbf{x}}       
\newcommand{\y}{\textbf{y}}       
\newcommand{\z}{\textbf{z}}

\renewcommand{\(}{\left (}
\renewcommand{\)}{\right )}

\newcounter{eqs}
\newcommand{\eql}[1]{\label{eq:#1} \tag{\arabic{eqs}} \stepcounter{eqs}}

\usepackage{arxiv}

\usepackage{graphicx}

\usepackage{doi}

\usepackage[toc,page]{appendix}

\usepackage{hyperref}
\usepackage{amsmath}
\usepackage{amsthm}

\theoremstyle{definition}

\newtheorem{theorem}{Theorem}

\newtheorem{lemma}[theorem]{Lemma}

\newtheorem{proposition}[theorem]{Proposition}

\newcommand{\handout}[5]{
   \renewcommand{\thepage}{#1-\arabic{page}}
   \noindent
   \begin{center}
   \framebox{
      \vbox{
    \hbox to 5.78in { {\bf ORIE 6334 Bridging Continuous and Discrete Optimization} \hfill #2 }
       \vspace{4mm}
       \hbox to 5.78in { {\Large \hfill #5  \hfill} }
       \vspace{2mm}
       \hbox to 5.78in { {\it #3 \hfill #4} }
      }
   }
   \end{center}
   \vspace*{4mm}
}

\usepackage{amsmath}
\usepackage{amsfonts}
\usepackage{amssymb}

\title{Integrating High-Dimensional Functions Deterministically}

\iftrue
\author{ 
{
	\hspace{1mm}David Gamarnik}
 \thanks{Massachusetts Institute of Technology, Sloan School of Management, Operations Research Center, Institute for Data, Systems and Society. Support from NSF grants CISE-2233897 and DMS-2015517 is gratefully acknowledged}  
 \And
{
\hspace{1mm}
Devin Smedira}\thanks{Massachusetts Institute of Technology, Operations Research Center} 
}

\fi
\date{}

\renewcommand{\headeright}{}
\renewcommand{\undertitle}{}
\renewcommand{\shorttitle}{Integrating High-Dimensional Functions Deterministically}

\hypersetup{
pdftitle={Temp},
pdfauthor={Devin ~Smedira, David~Gamarnik},
}

\usepackage{subfiles}
\begin{document}
\pagenumbering{roman}
\maketitle

\begin{abstract}
We design a Quasi-Polynomial time deterministic approximation algorithm for computing the integral of a multi-dimensional separable function, supported by some underlying hyper-graph structure, appropriately defined. Equivalently, our integral is the partition function of a graphical model with continuous potentials.  While randomized algorithms for high-dimensional integration are widely known, deterministic counterparts generally do not exist. We use the correlation decay method applied to the Riemann sum of the function to produce our algorithm. For our method to work, we require that the domain is bounded and the hyper-edge potentials are positive and bounded on the domain. We further assume that upper and lower bounds on the potentials separated by a multiplicative factor of $1 + O(1/\Delta^2)$, where $\Delta$ is the maximum degree of the graph. When $\Delta = 3$, our method works provided the upper and lower bounds are separated by a factor of at most $1.0479$. To the best of our knowledge, our algorithm is the first deterministic algorithm for high-dimensional integration of a continuous function, apart from the case of trivial product form distributions.
\end{abstract}

\newpage
\pagenumbering{arabic}



\section{Introduction}
In this paper, we study the problem of designing a deterministic approximation algorithm for computing  the integral of a multi-dimensional function over a bounded domain. In particular, given a function $f: \R^n \ra \R$, we are interested in computing an estimate of $\int_\D f d\lambda$ in deterministic polynomial in $n$ time, where $\lambda$ is the Lebesgue Measure and $\D$ is a rectangular domain.

The challenge of high-dimensional integration is a  central question across many scientific fields, including statistics and machine learning. For example, the question of conducting inference in graphical models, one of the cornerstone statistical models is reduced to the question of computing the associated partition function, which is a special case of function integration. 
Exact integration in high dimensions is known to be an intractable problem, specifically \#P-hard, even when  integrating functions as simple as indicators of a polytope (\cite{dyer1988complexity}). Thus, a longstanding problem has been to design algorithms to approximate high dimensional integrals. Traditional approaches rely on randomized methods to sample from the implied distribution, allowing one to then approximate the integrals true value. The Metropolis-Hastings algorithm, the development of which dates back to the 1950s (\cite{metropolis1953equation}, \cite{gelfand1990sampling}, \cite{hastings1970monte}), is one canonical example of this technique. Subsequent works have introduced several variations on the technique, such as the Metropolis-adjusted Langevin algorithm (\cite{roberts1996exponential}, the Hamiltonian Monte Carlo algorithm (\cite{duane1987hybrid}, \cite{Neal2012MCMC}), and the Metropolis-adjusted Proximal Algorithm (\cite{mou2022efficient}).

In this work, we take inspiration from recent breakthroughs in designing deterministic approximation algorithms for graph counting problems (
\cite{weitzCounting, BandyopadhyayGamarnikCounting, barvinok2017combinatorics})
for designing a deterministic algorithm for integration.
In graph counting problems, one is given a graph and a combinatorial structure (such as an independent set), and asked to count the number of such structures present in the graph. The exact forms of these problems are typically \#P-hard, and historically the only known approximations for them were randomized MCMC-based algorithms. However,  a new method called \textit{correlation decay} was introduced recently in  \cite{BandyopadhyayGamarnikCountingConference} and \cite{weitzCounting}, which allowed for deterministic approximation algorithms in graph counting problems. In particular, the algorithm in \cite{weitzCounting} is a Fully Polynomial Time Approximation Scheme, the strongest algorithm possible unless $P=NP$. Recent developments based on Spectral Independence have resulted in the improved exponent in the running time of the above algorithm which are now best possible (\cite{anari2021spectral}). Later, another deterministic method was introduced by Barvinok, which reformulates the problems in terms of partition function and analyzes the low order terms of a corresponding complex polynomial (\cite{barvinok2017combinatorics}). 

In loose terms, the correlation decay property states that when sampling a structure from the Gibbs distribution, the assignment given to vertices which are far apart in the graph are nearly independent. In particular, we require that the correlation between the two assignments decays exponentially as the distance increases. When this property holds, it becomes possible to estimate these assignment probabilities for each vertex by performing a recursive computation on a small neighborhood of the vertex. 

Despite the discrete nature of the property, and its initial use in solving discrete problems, there has been some recent work  applying the correlation decay method in the continuous setting. In particular, two recent works have studied polytopes associated with the linear programming relaxation of the independent set problem. In \cite{gamarnik2017uniqueness}, it was shown that when the underlying graph is a regular tree, the correlation decay property  holds. Subsequently, it was shown in \cite{gamarnik2023computing} that the property holds on all graphs, for a more restricted class of polytopes, allowing for the time first use of the correlation decay property for approximate computation of the  volume of a  polytope volume in quasi-polynomial time. Based on the apparent connection between the existence of randomized polynomial time algorithms (MCMC) and the correlation decay property, \cite{gamarnik2023computing} conjectured that the property should hold for \emph{any} polytope subject to minor regularity assumptions.

In this work, we turn instead to the question of integrating a general function, as opposed to the indicator functions arising in the context of computing the volume of a polytope.
We  now  discuss the results of our work and put them into context. Though we conjecture some form of the correlation decay property to hold for a wide class of integrable functions, it remains a challenge to define an appropriate notion of the correlation decay property for such a general setting, as the notions of ''graph distance'' and ''neighborhoods'' are not well defined. Thus, in this work we limit ourselves to studying functions $f:\R^n \ra \R$ which are defined with respect to some underlying  hyper-graph. In particular, we require that the function be a product of differentiable \textit{edge potential functions}, each of which depends only on the input values corresponding to the vertices in the edge. Equivalently we consider the graphical model setting with continuous potentials.
Under the assumption that the domain of the integration is a rectangle and the minimum and maximum value of the edge potential function over the relevant domain is within a multiplicative factor  $1 + O(1/(\eta \Delta))$ of each other, we design a (quasi)-polynomial deterministic approximation algorithm for the integral of $f$ over the associated domain. Here $\Delta$ is the maximum number of edges a vertex is a member of, and $\eta$ is the maximum number neighbors a vertex has in the underlying graph (notice that in the  graph setting, $\eta = \Delta$). When $\Delta=3$ this factor is $1.0479$ numerically.
While rather restrictive, we do believe it is of interest that non-trivial results can be achieved even under such restrictive assumptions for two reasons. First we don't adopt the standard assumptions which are adopted when analyzing MCMC type methods, 
such as convexity or log-Sobolev inequality, see more on this below.  In fact we believe that such inequality might be derivable from our correlation decay property as an implication. Second, we stress that our algorithm is detemirministic and thus first of a kind algorithm for integration outside of restricted case when integration can be done trivially, for example if the graphical model is product form.

Our algorithm is based on the correlation decay property  established for a discretized version of the integral problem, equivalent to the classical Riemann Sum. Our approach to prove the Correlation Decay property is to establish that the functions used to evaluate the marginal probabilities exhibits a contraction. A standard approach to achieve this was introduced in \cite{BandyopadhyayGamarnikCountingConference}, which we use as well, is to establish that the $\| \cdot \|_1$ of the gradient of the function is less than 1. We rely on the restriction on the values of the edge potential functions to prove the bound on the gradient norm. We then leverage this to design an algorithm which approximates a Riemann Sum of the function to within a factor of $1 + O(1/n)$ in time $n^{O(\log n))}$ using an algorithm similar to \cite{gamarnik2023computing}. We conjecture the result can be improved to be a true polynomial time algorithm by removing the dependence on $n$ in the exponent, as was done for many initial results based on the polynomial interpolation method (see \cite{patel2017deterministic}). Finally, we show that, so long as the edge potential functions have bounded gradients, the value of the Riemann sum well approximates the true value of the integral. Our approach is similar to the one introduced in~\cite{gamarnik2023computing}, but is in fact much simpler and more streamlined.

Comparing our results with the state of the art results in solving high-dimensional integration problems, such as \cite{mou2022efficient}, we note unlike the literature in this area, we do not assume that the function being integrated satisfies assumptions such as log-concavity, log-Sobolev or Poincare inequalities. Thus the existence of even randomized algorithms such as MCMC in our setting is not resolved. This explains why the multiplicative gap $1 + O(1/(\eta \Delta))$ appearing in our set of assumptions is so limited. We do conjecture that such algorithms can be constructed since the correlation decay property is usually ties with isoperimetric properties articulated by log-Sobolev and Poincare type bounds. We thus pose this as a conjecture: the Glauber dynamics (namely the canonical Markov chain with stationary distribution described by the function we integrate) mixes polynomially fast under the multiplicative gap assumption
$1 + O(1/(\eta \Delta))$. Similarly to~\cite{gamarnik2023computing} we furthermore believe that these assumptions can be significantly relaxed. In general, we paraphrase
this as the scientific challenge of establishing deterministic counterparts
to randomized integration scheme such as MCMC or Metropolis-Hastings. The $P=RP$ and $P=BPP$ conjectures provide the basis to believe that such deterministic schemes should exist.

We conclude with a brief overview of the following sections.
In section \ref{definitions} we formally introduce the problem and state the main theorem. Section \ref{preliminary} contains some preliminary results 
related to the recursive problem construction. 
Section \ref{correlation} will be where we derive the key Correlation Decay Property, which we use to design our algorithm in section \ref{algorithm}. Finally, in section \ref{disc to cont}, we will relate the discrete and continuous versions of our problem and prove the main theorem.

\section{Model and the results} \label{definitions}
We now formally introduce the framework presented in the introduction, as well as the main theorem. Consider a hypergraph $G = (V,E)$ with $n$ vertices of some fixed maximum degree $\Delta$ (meaning each vertex in the hypergraph is a member of at most $\Delta$ edges). We will further assume that each vertex has at most $\eta$ neighbors, where a neighbor is defined as any other vertex which shares at least one edge (notice that $\eta \leq R \Delta$). We will define a \textit{hypergraph potential function} to be any function $f(\y):\R^{n} \ra \R$ which takes the form
\[f(\y) = \prod_{e \in E} \Phi_e(\y_e). \eql{f def} 
\]
We refer to each function $\Phi_e$ as the edge potential function for the edge $e$. We wish to compute the expression
\[\int 1(\y \in [0,1]^{n}) f(\y) d\lambda,\]
where $\lambda$ is the Lebesgue measure. Though we define the region of integration to be $[0,1]^n$ here for convenience, similar results will hold for any rectangular domain. Rather than computing a direct approximation of this integral, we will instead compute an approximation of the corresponding Riemann Sum. Fix some relaxation parameter $N$, and let $\D = \{\y/N| \y \in \Z^n, 0 \leq y_v < N \forall v \in V\}$. We define the partition function
\[Z(G) = \sum\limits_{\y \in \D} f(\y), \eql{Z def} 
\]
which is equivalent to a traditional Riemann Sum scaled up by a factor of $N^n$, with the goal of approximating this value. For the remainder of this paper, we will restrict ourselves to the setting where each $\Phi_e$ above satisfies 
\[c_e \leq \Phi_e(\y_e) \leq (1 - \delta)^{1/2\Delta} c_e\(1 + \frac{\log(1 + \frac{1}{\eta})}{2\Delta}\),
\eql{Phi assum} 
\]
for some constant $c_e > 0$ unique to each edge and a universal constant $\delta > 0$. In this setting, we will be able to show that the \textit{correlation decay} property holds, allowing us to approximate the value $Z(G)$. We will then be able to relate this approximation back to the value of the original integral. In particular, we will show the following.

\begin{theorem}[Main Result] \label{main result}
Suppose we have a function $f$ taking the form of equation \ref{eq:f def} with respect to some hypergraph $G$. Further, suppose each function $\Phi_e$ is differentiable, satisfies the condition in equation \ref{eq:Phi assum}, and there exists some $k > 0$ such that every $\Phi_e$ satisfies $\|\nabla \Phi_e(\y_e)\|_2 \leq k c_e$ for every $\y \in [0,1]^n$. Then, there exists an algorithm to produce an estimate $\tilde V$ of $\int 1(\y \in [0,1]^{n}) f(\y) d\lambda$ such that
\[\bigg | \frac{\tilde V}{\int 1(\y \in [0,1]^{n}) f(\y) d\lambda} - 1 \bigg | \leq O(1/n).\]
Taking $k, \Delta,$ and $ \eta$ to be constants, this algorithm will run in $n^{O(\log(n))}$ time. 
\end{theorem}

In the course of our approximation algorithm, we will need to iteratively restrict the domain of our partition function. Thus, we will introduce constraints of the form $v \la m$, equivalent to restricting $y_v = m$, and let $\beta$ be collections of such constraints. We then define $\D_\beta \subset \D$ to be the set of points in $\D$ that satisfy every constraint in $\beta$. We will use the notation $(\beta, v \la m)$ to denote $\beta \cup \{v \la m\}$. With this notation, we can define the following family of restricted partition functions
\[Z(G, \beta) =  \sum\limits_{\y \in \D_\beta} f(\y). 
\eql{Z beta} 
\]
We can then use this to define a probability distribution over the assignments to each vertex in the hypergraph $G$ induced by the function $f$ and constraints $\beta$. In particular, we define
\[x(G, v \la m, \beta) = \frac{Z(G,(\beta, v \la m))}{Z(G,\beta)},\]
which for each well formulated $\beta$ will define a probability distribution over the assignments 0 to $N-1$ for the vertex $v$. Our algorithm will rely on computing an approximation $\tilde x(G, v \la m, \beta)$ of the above value, which will be done by exploiting the recursive nature of the problem. In particular, if we order the vertices 1 through $n$ and let $\beta_i = (v_j \la 0, j \leq i)$ for $i \in [0,n]$, with $\beta_0 = \{\}$, we can see
\begin{align*}
Z(G) &= x^{-1}(G, v_1 \la 0, \beta_0 ) Z(G, \beta_1)\\
&= \prod\limits_{i = 1}^{n} x^{-1}(G, v_i \la 0, \beta_{i-1}) Z(G, \beta_n)\\
&= f(\0) \prod\limits_{i = 1}^{n} x^{-1}(G, v_i \la 0, \beta_{i-1}),
\eql{Z-unfurl} 
\end{align*}
since $Z(G, \beta_n) = f(\0)$ by construction. Thus, if we can approximate the values of $x(G, v_i \la m, \beta_{i-1})$ within a factor of $\frac{1}{n^2}$, we will be able to approximate $Z(G)$ within a factor of $\frac{1}{n}$. We will formalize this notion in section \ref{algorithm}.

\section{Preliminary Results} \label{preliminary}

Fix a vertex $v \in V$ and a set of constraints $\beta$. Let $v_1, \ldots v_{\eta'}$, $\eta' \leq \eta$, be the unconstrained neighbors of $v$. For any sequence $y_1, \ldots y_{\eta'} \in [0,N-1]$ and any $i \in [1,\eta']$, we will use $\bar y_i$ to denote $(y_1, \ldots y_i)$. For each such sequence, we can consider the associated marginal probabilities
\[x(\bar y_i) \triangleq x(G \setminus v, v_i \la y_i, (\beta, (v_j \la y_j, j < i))). \]
We will use $\x$ to denote the vector of marginals $(x(\bar y_i), 1 \leq i \leq \eta', \bar y_i \in[0,N-1]^i)$. Notice that for any fixed $i$ and $\bar y_{i-1}$, we have
\[\sum\limits_{y_i=0}^{N-1} x(\bar y_i) = \frac{\sum\limits_{y_i=0}^{N-1} Z(G \setminus v, (\beta, (v_j \la y_j, j \leq i)))}{Z(G \setminus v, (\beta, (v_j \la y_j, j < i)))} =\frac{Z(G \setminus v, (\beta, (v_j \la y_j, j < i)))}{Z(G \setminus v, (\beta, (v_j \la y_j, j < i)))} = 1, \]
which inductively implies
\[
\sum\limits_{y_1, \ldots y_i \in [0,N-1]} \prod\limits_{j=1}^i x(\y_i) = 1.
\eql{1-sum} 
\]
We now define the function
\[g_m(\x) = 
\frac{\sum\limits_{y_1, \ldots y_{\eta'} \in [0,N-1]} 
\prod\limits_{j=1}^{\eta'} x(\bar y_j)
\prod\limits_{e \in E, v \in e} \Phi_{e}(m, \y_{e\setminus v})}
{\sum\limits_{y_0, y_1, \ldots y_{\eta'} \in [0,N-1]} 
\prod\limits_{j=1}^{\eta'} x(\bar y_j)
\prod\limits_{e \in E, v \in e} \Phi_{e}(\y_e)}
\eql{g_m} 
\]
for every $m \in [0,N-1]$. This function captures the recursive structure of the problem, as we will see when we use the following proposition in the next section.
\begin{proposition}
\label{recursion}
The following relation holds for every 
$m \in [0,N-1]$ and every $\beta$:
\begin{align*}
x(G,v \leftarrow m, \beta)=g_m(\x).
\end{align*}
\end{proposition}

\begin{proof}
Starting from equation \ref{eq:Z beta}, we see that
\begin{align*}
Z(G,\beta) &= \sum\limits_{\y \in D_\beta} f(\y)\\
&= \sum\limits_{\y \in D_\beta} \prod\limits_{e \in E} \Phi_{e}(\y_e)\\
&= \sum\limits_{y_0, y_1, \ldots y_{\eta'} \in [0,N-1]} \prod_{e \in E, v \in e} \Phi_{e}(\y_e) \sum\limits_{\y\setminus\{0,\ldots \eta'\} \in D_{\beta}} \prod\limits_{e \in E, v \in e} \Phi_{e}(\y_e)\\
&= \sum\limits_{y_0, y_1, \ldots y_{\eta'} \in [0,N-1]} \prod_{e \in E, v \in e} \Phi_{e}(\y_e) Z(G\setminus v, (\beta, (v_i \la y_i, i \in 1, \ldots \eta')))\\
&= \sum\limits_{y_0, y_1, \ldots y_{\eta'} \in [0,N-1]} 
\frac{\prod\limits_{j=1}^{\eta'} Z(G \setminus v, (\beta, (v_i \la y_i, i \leq j)))}
{\prod\limits_{j=1}^{\eta'-1} Z(G \setminus v, (\beta, (v_i \la y_i, i \leq j)))}
\prod_{e \in E, v \in e} \Phi_{e}(\y_e) \\
&= \sum\limits_{y_0, y_1, \ldots y_{\eta'} \in [0,N-1]} 
Z(G \setminus v, (\beta, v_1 \la y_1))
\prod\limits_{j=2}^{\eta'} x(\bar y_j)
\prod_{e \in E, v \in e} \Phi_{e}(\y_e). \\
\end{align*}
By a similar line of reasoning we will also get that 
\[Z(G, (\beta, v \la m)) = \sum\limits_{ y_1, \ldots y_{\eta'} \in [0,N-1]} 
Z(G \setminus v, (\beta, v_1 \la y_1))
\prod\limits_{j=2}^{\eta'} x(\bar y_j)
\prod_{e \in E, v \in e} \Phi_{e}(m, \y_{e\setminus v}).\]
Combining these, we can see that
\begin{align*}
x(G, v \la m, \beta) &= \frac{Z(G, (\beta, v \la m))}{Z(G, \beta)}
\\&=
\frac{\sum\limits_{y_1, \ldots y_{\eta'} \in [0,N-1]} 
Z(G \setminus v, (\beta, v_1 \la y_1))
\prod\limits_{j=2}^{\eta'} x(\bar y_j)
\prod_{e \in E, v \in e} \Phi_{e}(m, \y_{e\setminus v})}
{\sum\limits_{y_0, y_1, \ldots y_{\eta'} \in [0,N-1]} 
Z(G \setminus v, (\beta, v_1 \la y_1))
\prod\limits_{j=2}^{\eta'} x(\bar y_j)
\prod_{e \in E, v \in e} \Phi_{e}(\y_e)}\\
&= 
\frac{\sum\limits_{y_1, \ldots y_{\eta'} \in [0,N-1]} 
\prod\limits_{j=1}^{\eta'} x(\bar y_j)
\prod_{e \in E, v \in e} \Phi_{e}(m, \y_{e\setminus v})}
{\sum\limits_{y_0, y_1, \ldots y_{\eta'} \in [0,N-1]} 
\prod\limits_{j=1}^{\eta'} x(\bar y_j)
\prod_{e \in E, v \in e} \Phi_{e}(\y_e)},
\end{align*}
where the last equality holds by dividing the numerator and denominator by $Z(G\setminus v,\beta)$. This completes the proof.
\end{proof}

\section{Correlation Decay} \label{correlation}
This section will be dedicated to establishing the necessary correlation decay property, to be used later in our final algorithm.

\subsection{Computation of Derivatives}

In this next section, we will compute bounds on the values of the derivatives of our function $g_m$. For the purpose of clarity, we define two expression to represent the numerator and denominator of our function $g_m$,
\begin{align*}
U_m &= \sum\limits_{y_1, \ldots y_{\eta'} \in [0,N-1]} 
\prod\limits_{j=1}^{\eta'} x(\bar y_j)
\prod_{e \in E, v \in e} \Phi_{e}(n, \y_{e\setminus v})\\
L = \sum\limits_{l = 0}^{N-1} U_l &=  
\sum\limits_{l=0}^{N-1} \sum\limits_{y_1, \ldots y_{\eta'} \in [0,N-1]} 
\prod\limits_{j=1}^{\eta'} x(\bar y_j)
\prod_{e \in E, v \in e} \Phi_{e}(l, \y_{e\setminus v}).
\end{align*}
First, notice that for any value of $\x$ we have that
\[L \geq \sum\limits_{l=0}^{N-1} \sum\limits_{y_1, \ldots y_{\eta'} \in [0,N-1]} 
\prod\limits_{j=1}^{\eta'} x(\bar y_j)
\prod_{e \in E, v \in e} c_e 
= N \prod_{e \in E, v \in e} c_e, 
\eql{L Bound}\] 
by equations \ref{eq:Phi assum} and \ref{eq:1-sum}. Now, fix some $k \in [1, \eta']$ and some vector $\bar y_k \in [0,N-1]^k$. We have
\begin{align*}
\frac{\partial U_m}{\partial x(\bar y_k)} &= \prod\limits_{j=1}^{k-1} x(\bar y_j) \sum\limits_{y_{k+1}, \ldots y_{\eta'} \in [0,N-1]} 
\prod\limits_{j=k+1}^{\eta'} x(\bar y_j)
\prod_{e \in E, v \in e} \Phi_{e}(m, \y_{e\setminus v})
\\
\frac{\partial L}{\partial x(\bar y_k)} = \sum\limits_{l=0}^{N-1} \frac{\partial U_l}{\partial x(\bar y_k)}
&= \sum\limits_{l=0}^{N-1} \prod\limits_{j=1}^{k-1} x(\bar y_j)
\sum\limits_{y_{k+1}, \ldots y_{\eta'} \in [0,N-1]} 
\prod\limits_{j=k+1}^{\eta'} x(\bar y_j)
\prod_{e \in E, v \in e} \Phi_{e}(l, \y_{e\setminus v}).
\end{align*}
Using these definitions, we compute

\begin{align*}
&\left | L \frac{\partial U_m}{\partial x(\bar y_k)} - U_m \frac{\partial L}{\partial x(\bar y_k)} \right | x(\bar y_k) \\
=& \Bigg | \left [
\sum\limits_{l=0}^{N-1} \sum\limits_{z_1, \ldots z_{\eta'} =0}^{N-1} 
\prod\limits_{j=1}^{\eta'} x(\bar z_j)
\prod_{e \in E, v \in e} \Phi_{e}(l, \z_{e\setminus v})
\right ] \left [
\sum\limits_{y_{k+1}, \ldots y_{\eta'} = 0}^{N-1} 
\prod\limits_{j=1}^{\eta'} x(\bar y_j)
\prod_{e \in E, v \in e} \Phi_{e}(m, \y_{e\setminus v})
\right ]\\
-& \left [
\sum\limits_{z_1, \ldots z_{\eta'} = 0}^{N-1} 
\prod\limits_{j=1}^{\eta'} x(\bar z_j)
\prod_{e \in E, v \in e} \Phi_{e}(m, \z_{e\setminus v})
\right ] \left [
\sum\limits_{l=0}^{N-1} \sum\limits_{y_{k+1}, \ldots y_{\eta'} = 0}^{N-1} 
\prod\limits_{j=1}^{\eta'} x(\bar y_j)
\prod_{e \in E, v \in e} \Phi_{e}(l, \y_{e\setminus v})
\right ]  \Bigg |\\
\leq& \sum\limits_{l=0}^{N-1} \sum\limits_{z_1, \ldots z_{\eta'} =0}^{N-1}  \sum\limits_{y_{k+1}, \ldots y_{\eta'} = 0}^{N-1} 
\prod\limits_{j=1}^{\eta'} x(\bar z_j) x(\bar y_j)\\
& \left | \prod_{e \in E, v \in e} \Phi_{e}(l, \z_{e\setminus v}) \Phi_{e}(m, \y_{e\setminus v}) -  \prod_{e \in E, v \in e} \Phi_{e}(m, \z_{e\setminus v}) \Phi_{e}(l, \y_{e\setminus v})
\right |\\
\leq& \sum\limits_{l=0}^{N-1} \sum\limits_{z_1, \ldots z_{\eta'} = 0}^{N-1}  \sum\limits_{y_{k+1}, \ldots y_{\eta'} = 0}^{N-1} 
\prod\limits_{j=1}^{\eta'} x(\bar z_j) x(\bar y_j)
\(\prod\limits_{e \in E, v \in E} c_e^2 \) \((1 - \delta)\(1 + \frac{\log(1 + \frac{1}{\eta})}{2 \Delta}\)^{2\Delta} - 1 \)\\
\leq& \sum\limits_{l=0}^{N-1} \sum\limits_{z_1, \ldots z_{\eta'} = 0}^{N-1}  \sum\limits_{y_{k+1}, \ldots y_{\eta'} = 0}^{N-1} 
\prod\limits_{j=1}^{\eta'} x(\bar z_j) x(\bar y_j)
\(\prod\limits_{e \in E, v \in E} c_e^2 \) (1 - \delta) \(\(1 + \frac{\log(1 + \frac{1}{\eta})}{2 \Delta}\)^{2\Delta} - 1 \)\\
\leq& (1 - \delta) \(e^{\log(1 + 1/\eta)} - 1 \) \sum\limits_{m=0}^{N-1} \sum\limits_{z_1, \ldots z_{\eta'} = 0}^{N-1}  \sum\limits_{y_{k+1}, \ldots y_{\eta'} = 0}^{N-1} 
\prod\limits_{j=1}^{\eta'} x(\bar z_j) x(\bar y_j)
\prod\limits_{e \in E, v \in e} c_e^2 \\
=& \frac{1-\delta}{\eta}  \sum\limits_{m=0}^{N-1} \sum\limits_{z_1, \ldots z_{\eta'} = 0}^{N-1}  \sum\limits_{y_{k+1}, \ldots y_{\eta'} = 0}^{N-1} 
\prod\limits_{j=1}^{\eta'} x(\bar z_j) x(\bar y_j)
\prod\limits_{e \in E, v \in e} c_e^2 \\
=& \frac{N(1-\delta)}{\eta} \prod\limits_{j=1}^{k} x(\bar y_j) \prod\limits_{e \in E, v \in E} c_e^2,
\end{align*}
where the second inequality comes from equation \ref{eq:Phi assum} and the last equality comes from equation \ref{eq:1-sum}. Combining this with equation \ref{eq:L Bound}, we get
\[\left |\frac{\partial g_m(\x)}{\partial x(\bar y_k)} \right | = 
\left | \frac{L \frac{\partial U_n}{\partial x(\bar y_k)} - \frac{\partial U_n}{\partial x(\bar y_k)} L}{L^2} \right | \leq
\frac{\frac{N(1-\delta)}{\eta} \prod\limits_{j=1}^{k-1} x(\bar y_j) \prod\limits_{e \in E, v \in E} c_e^2}
{\(N \prod_{e \in E, v \in e} c_e\)^2}
= \frac{(1-\delta) \prod\limits_{j=1}^{k-1} x(\bar y_j)}{\eta N}.
\eql{derivative} 
\]

\subsection{Bounding the Gradient}

The remainder of this section will be dedicated to proving the following theorem.
\begin{theorem} [Bounded Gradients]
\label{Bounded Gradients}
For any $m \in [0,N-1]$, we have that
\[\|\nabla g_m \|_1 \leq 1-\delta.\]
\end{theorem}
\begin{proof}
Fix some $k \in [1,\eta']$. Using equation \ref{eq:derivative}, we define
\[p_k \triangleq 
\sum\limits_{\bar y_k \in [0,N-1]^k} \left |\frac{\partial g_m(\x)}{\partial x(\bar y_k)} \right |,
\]
and use equations \ref{eq:derivative} and \ref{eq:1-sum} to see
\begin{align*}
p_k \leq \sum\limits_{\bar y_k \in [0,N-1]^k} \frac{(1-\delta)\prod\limits_{j=1}^{k-1} x(\bar y_j)}{\eta N} 
= \sum\limits_{y_k=0}^{N-1} \frac{1-\delta}{\eta N} \sum\limits_{\bar y_{k-1} \in [0,N-1]^{k-1}} \prod\limits_{j=1}^{k-1} x(\bar y_j) 
= \frac{1-\delta}{\eta}.
\end{align*}
Thus, since the vertex $v$ has $\eta' < \eta$ neighbors, we get that
\[\|\nabla g_m \|_1 = \sum\limits_{k=1}^{\eta'} p_k \leq  \sum\limits_{k=1}^\eta \frac{1-\delta}{\eta} = 1-\delta. \]
\end{proof}

\section{The Approximation Algorithm} \label{algorithm}
This section will be dedicated to the algorithm used to approximate the value $Z(G)$. The primary piece in this algorithm, present in algorithm \ref{alg:prob}, will be a recursive algorithm to approximate the value of $x(G, v \la m,\beta)$.

\SetKwComment{Comment}{\# }{}
\begin{algorithm}[ht]
\caption{ApproxProb}\label{alg:prob}
\KwData{Hypergraph $G$,  Constraints $\beta$, Recursion Depth $d$, Assignment $y$}
\KwResult{Estimate $\tilde x(G, v \la m, \beta)$ of $x(G, v \la m, \beta)$}
\lIf{$d=0$}{\Return $\frac{1}{N}$}
Let $v_1, \ldots v_{\eta'}$ be the unconstrained neighbors of $v$\;
\For{$j \in [1,{\eta'}]$}{
\For{$y_1, \ldots y_j \in [0,N-1]$}{
Let $\tilde x(\bar y) = \operatorname{ApproxProb}(G \setminus v, (\beta, (v_i \la y_i, i < j)), d-1, y_j)$\;
}
}
\Return{$g_m(\tilde \x)$} \Comment*[l]{where $g_m$ is defined in equation \ref{eq:g_m}}
\end{algorithm}
\begin{lemma}
\label{prob bound}
Algorithm \ref{alg:prob} will produce an estimate $\tilde x(G, v \la m, \beta)$ satisfying
\[|\tilde x(G, v \la m, \beta) - x(G, v \la m, \beta)| \leq (1 - \delta)^d.\]
This estimate will be computed in time $O(N^{\eta d})$.
\end{lemma}
\begin{proof}
We will prove the above theorem by induction. When $d=0$, the difference is trivially bounded by 1, so the theorem will hold. For $d > 1$, we have by induction that the coordinates of $\tilde \x$ computed in the highest level run of the algorithm differ from the coordinates of $\x$ by at most $(1-\delta)^{d-1}$. Further, by the mean value theorem, there must be some linear interpolation $\hat \x$ of $\tilde \x$ and $\x$ which satisfies
\[g_m(\tilde \x) - g_m(\x) = \nabla g_m(\hat \x) \cdot (\tilde \x - \x).\]
Thus, we can use theorem \ref{Bounded Gradients} to conclude that
\[|g_m(\tilde \x) - g_m(\x)| \leq \|\nabla g_m(\hat \x)\|_1 \|(\tilde \x - \x)\|_\infty \leq (1-\delta)^d.\]

Additionally, we wish to analyze the runtime of algorithm \ref{alg:prob}. We will let $T(d)$ denote the runtime of the algorithm with recursion depth $d$, and observe that $T(0) = O(1)$. For higher values of $d$, the limiting step will be the runtime of the recursive calls. There will be exactly $N^{\eta'}$ such calls, and each one will take at most $O({\eta'})$ time to update the value of $\beta$. Thus, we get that $T(d) = N^{\eta'} O({\eta'}) T(d-1)$, implying that $T(d) = O(N^{\eta d})$, since ${\eta'} \leq \eta$. 
\end{proof}

Before we analyze the full algorithm, we will make an observation about the function $g_m$. In particular, we will require the following Lemma.

\begin{lemma}
\label{g_m lower lemma}
For any vector of marginal probabilities $\x$, we have
\[g_m(\x) \geq \frac{1}{\sqrt{2}N},\]
as long as the underlying hypergraph $G$ is connected.
\end{lemma}
\begin{proof}
So long as the input vector $\x$ forms a probability distribution, we use equation \ref{eq:g_m} to see
\begin{align*}
g_m(\x) &= 
\frac{\sum\limits_{y_1, \ldots y_{\eta'} \in [0,N-1]} 
\prod\limits_{j=1}^{\eta'} x(\bar y_j)
\prod\limits_{e \in E, v \in e} \Phi_{e}(m, \y_{e\setminus v})}
{\sum\limits_{y_0, y_1, \ldots y_{\eta'} \in [0,N-1]} 
\prod\limits_{j=1}^{\eta'} x(\bar y_j)
\prod\limits_{e \in E, v \in e} \Phi_{e}(\y_e)}\\
&\geq 
\frac{\sum\limits_{y_1, \ldots y_{\eta'} \in [0,N-1]} 
\prod\limits_{j=1}^{\eta'} x(\bar y_j)
\prod\limits_{e \in E, v \in e} c_e}
{\sum\limits_{y_0, y_1, \ldots y_{\eta'} \in [0,N-1]} 
\prod\limits_{j=1}^{\eta'} x(\bar y_j)
\prod\limits_{e \in E, v \in e} c_e \(1 + \frac{\log(1 + \frac{1}{\eta})}{2 \Delta}\)}\\
&\geq 
\frac{\sum\limits_{y_1, \ldots y_{\eta'} \in [0,N-1]} 
\prod\limits_{j=1}^{\eta'} x(\bar y_j)}
{\(1 + \frac{\log(1 + \frac{1}{\eta})}{2 \Delta}\)^\Delta
\sum\limits_{y_0, y_1, \ldots y_{\eta'} \in [0,N-1]} 
\prod\limits_{j=1}^{\eta'} x(\bar y_j)
 }\\
 &= \frac{1}{N \(1 + \frac{\log(1 + \frac{1}{\eta})}{2 \Delta}\)^\Delta},
\end{align*}
where the first inequality follows from the assumption in equation \ref{eq:Phi assum}, the second inequality follows from $v$ being a member of at most $\Delta$ edges, and the final equality follows from equation \ref{eq:1-sum}. Further, we can observe that
\[\(1 + \frac{\log(1 + \frac{1}{\eta})}{2 \Delta}\)^\Delta \leq \exp\(\log\(1 + \frac{1}{\eta}\)/2\) = \sqrt{1 + \frac{1}{\eta}} \leq \sqrt{2}, \]
so long as the hypergraph is connected. Thus,
\[g_m(\x) \geq \frac{1}{\sqrt{2}N}.
\]
\end{proof}
With this Lemma, we conclude with the following theorem.

\begin{algorithm}[ht]
\caption{ApproxZ}\label{alg:Z}
\KwData{Hypergraph $G$, Recursion Depth $d$}
\KwResult{Estimate $\tilde Z(G)$ of $Z(G)$}
Let $z\_inv \la 1$ and $\beta \la \{\}$\;
\For{$v \in V$}{
$z\_inv \la z\_inv * \rm ApproxProb(G, \beta, d, 0)$\;
$\beta \la (\beta, v \la 0)$\;
}
\Return{$f(\0)/ z\_inv$}
\end{algorithm}
\begin{theorem}
\label{final alg}
The estimate $\tilde Z(G)$ of $Z(G)$ produced by algorithm \ref{alg:Z} will satisfy
\[1 - O(1/n) \leq \frac{\tilde Z(G)}{Z(G)} \leq 1 + O(1/n)\]
for some $d = O(\log(Nn))$. This estimate will be computed in time $N^{O(\eta \log( N n))}$.
\end{theorem}
\begin{proof}
Order the vertices $v_1$ to $v_n$ in the same order they are iterated in algorithm \ref{alg:Z}, and let $\beta_i = (v_j \la 0, j \leq i)$ as in the end of section \ref{definitions}. Set $d = \log_{1 - \delta}(\frac{1}{\sqrt{2}Nn^2})$. This will result in our approximations $\tilde x(G, v_i \la 0, \beta_{i-1})$ computed in the algorithm to satisfy
\[| x(G, v_i \la 0, \beta_{i-1}) - \tilde x(G, v_i \la 0, \beta_{i-1})|
\leq (1 - \delta)^d = \frac{1}{\sqrt{2}Nn^2} \leq \frac{\tilde x(G, v_i \la 0, \beta_{i-1})}{n^2}\]
by Lemma \ref{prob bound} and \ref{g_m lower lemma}. Dividing both sides by $\tilde x(G, v_i \la 0, \beta_{i-1})$, we get
\[\bigg |  \frac{ x(G, v_i \la 0, \beta_{i-1})}{\tilde x(G, v_i \la 0, \beta_{i-1})} - 1 \bigg | \leq \frac{1}{n^2}.
\]
Next, notice that 
\[\frac{\tilde Z(G)}{Z(G)} = \frac{f(\0) \prod\limits_{i = 1}^{n} \tilde x^{-1}(G, v_i \la 0)}{f(\0) \prod\limits_{i = 1}^{n} x^{-1}(G, v_i \la 0)} = \frac{\prod\limits_{i = 1}^{n} x(G, v_i \la 0)}{\prod\limits_{i = 1}^{n} \tilde x(G, v_i \la 0)},\]
by equation \ref{eq:Z-unfurl}. Therefore, the estimate $\tilde Z$ produced by algorithm \ref{alg:Z} will satisfy
\[\(1- \frac{1}{n^2}\)^n = 1 - O(1/n) \leq \frac{\tilde Z(G)}{Z(G)} \leq 1 + O(1/n) \leq \(1 + \frac{1}{n^2}\)^n.\]
The runtime of this algorithm will be dominated by the calls to the function ApproxProb, which each take time $O(N^{\eta d})$ by Lemma \ref{prob bound}. Further, notice that $d = O(\log(Nn))$ since $(1 - \delta) < 1$. Thus, since there are $n$ calls to this function with this $d$, the algorithm will take time $N^{O(\eta \log( Nn))}$.
\end{proof}

\section{From Discrete to Continuous} \label{disc to cont}
In this section, we will present one method for relating the value of a Riemann sum to the value of the corresponding integral, based on bounding the functions $l2$-norm. This process will conclude with a short proof of theorem \ref{main result}. 

\begin{lemma}
\label{integral approx}
Suppose we have a function $f$ defined as in equation \ref{eq:f def}, where each function $\Phi_e$ is differentiable and satisfies $\|\nabla \Phi_e(\x)\|_2 \leq kc_e$ for every $\x \in [0,1]^n$. Then,
\[\( 1 - \frac{\sqrt{n}k}{N} \)^{|E|} 
\leq \frac{\int 1(\y \in [0,1]^n) f(\y) d\lambda}{Z(G)/N^n}
\leq \( 1 + \frac{\sqrt{n}k}{N} \)^{|E|}
,\]
where $Z(G)$ is defined as in equation \ref{eq:Z def}.
\end{lemma}

\begin{proof}
Pick any vector $\y \in \D$, and let $\y'$ be any vector satisfying $y_i \leq y_i' < y_i + 1/N$ for each index $i$. We will denote the set of all such $\y'$ as $\D_\y$. Notice that $\|\y - \y'\|_2 \leq \frac{\sqrt{n}}{N}$. Thus, for every edge $e \in E$, we have 
\[\Phi_e(\y) - \frac{\sqrt{n}}{N} k c_e \leq \Phi_e(\y') \leq \Phi_e(\y) + \frac{\sqrt{n}}{N}kc_e. \]
Using this, we see
\begin{align*}
\frac{\int_{\D_\y} f(\x) d\x}{f(\y)/N^n}
\leq \frac {\int_{\D_\y} \prod\limits_{e \in E} \(\Phi_e(\y) + \frac{\sqrt{n}}{N}kc_e d\x \)}{\prod\limits_{e \in E} \Phi_e(\y)/N^n}
= \frac{\prod\limits_{e \in E}\(\Phi_e(\y) + \frac{\sqrt{n}}{N}kc_e\)}{\prod\limits_{e\in E} \Phi_e(\y)}
\leq \(
1 + \frac{\sqrt{n}k}{N}
\)^{|E|},
\end{align*}
where the first inequality follows from the above equation, and the second inequality follows from the assumption in equation \ref{eq:Phi assum}. Similar reasoning will yield that 
$\frac{\int_{\D_\y} f(\x) d\x}{f(\y)/N^n} \geq \bigg (
1 - \frac{\sqrt{n}k}{N}
\bigg )^{|E|}$.
Finally, we observe that
\[\frac{\int 1(\y \in [0,1]^n) f(\y) d\lambda}{Z(G)/N^n} = \frac{\sum\limits_{\y \in \D} \int_{\D_\y} f(\x) d\x}{\sum\limits_{\y \in \D} f(\y)/N^n} \leq \max_{\y \in \D} \frac{\int_{\D_\y} f(\x) d\x}{f(\y)/N^n} \leq \(
1 + \frac{\sqrt{n}k}{N}
\)^{|E|},\]
with similar logic yielding
\[\frac{\int 1(\y \in [0,1]^n) f(\y) d\lambda}{Z(G)/N^n} \geq \( 1 - \frac{\sqrt{n}k}{N} \)^{|E|}.\]
\end{proof}
We are now ready to prove our main result.

\textbf{Proof of Theorem \ref{main result}}\\
Let $N = k n^{3/2} |E|$. By Lemma \ref{integral approx}, we will have that
\[e^{-1/n} \leq \frac{\int 1(\y \in [0,1]^n) f(\y) d\lambda}{Z(G)/N^n} \leq e^{1/n},\]
implying that
\[1 - O(1/n) \leq \frac{Z(G)/N^n}{\int 1(\y \in [0,1]^n) f(\y) d\lambda} \leq 1 + O(1/n).\]
Now, let $\tilde Z(G)$ be the estimate of $Z(G)$ produced by algorithm \ref{alg:Z} when $d = O(\log(Nn)) = O(\log(n k |E|))$. By theorem \ref{final alg}, we must have that
\[1 - O(1/n) \leq \frac{\tilde Z(G)}{Z(G)} \leq 1 + O(1/n).\]
Finally, we observe that
\[\(\frac{Z(G)/N^n}{\int 1(\y \in [0,1]^n) f(\y) d\lambda}\)\(\frac{\tilde Z(G)}{Z(G)}\) = \frac{\tilde Z(G)/N^n}{\int 1(\y \in [0,1]^n) f(\y) d\lambda},  \]
which means
\[(1-O(1/n))^2 = 1 - O(1/n) \leq \frac{\tilde Z(G)/N^n}{\int 1(\y \in [0,1]^n) f(\y) d\lambda} \leq 1 + O(1/n) = (1 + O(1/n))^2.\]
Thus, if we set our estimate of the integral $\tilde V = \tilde Z(G)/N^n$, we will have that
\[\bigg | \frac{\tilde V}{\int 1(\y \in [0,1]^n) f(\y) d\lambda} - 1 \bigg | \leq O(1/n).\]
Further, the runtime to compute this estimate will be dominated by the runtime of algorithm \ref{alg:Z}. By Theorem \ref{final alg}, this will be $(k n |E|)^{O( \eta \log( k n |E|))}$. Observing $|E| \leq \Delta n$ and taking $\eta$, $k$, and $\Delta$ to be constants, we get a runtime of $n^{O(\log n)}$.

\bibliography{main}







\end{document}